\documentclass[10pt,letterpaper]{article}

\usepackage{cogsci}
\usepackage{amsmath,amssymb}
\usepackage{commath}
\usepackage{booktabs}

\cogscifinalcopy 

\usepackage[
    backend=biber,
    style=apa,
    natbib=true,
    doi=false,
    isbn=false,
    url=false,
]{biblatex}
\usepackage{pslatex}
\usepackage{float} 

\newcommand\blfootnote[1]{%
  \begingroup
  \renewcommand\thefootnote{}\footnote{#1}%
  \addtocounter{footnote}{-1}%
  \endgroup
}
\usepackage{graphicx}
\graphicspath{ {./images/} }

\title{Adaptive Social Learning using Theory of Mind}
 
\author{\\\\{\large \bf Lance Ying$^{1,2}$, Ryan Truong$^{1}$, Joshua B. Tenenbaum$^{2}$, Samuel J. Gershman$^{1}$} \\ \\
  $^{1}$Harvard University, Cambridge, MA, USA\\
  $^{2}$Massachusetts Institute of Technology, Cambridge, MA, USA  \\
  }

\begin{document}

\maketitle

\begin{abstract}
Social learning is a powerful mechanism through which agents learn about the world from others. However, humans don't always choose to observe others, since social learning can carry time and cognitive resource costs. How do people balance social and non-social learning? In this paper, we propose a rational mentalizing model of the decision to engage in social learning. This model estimates the utility of social learning by reasoning about the other agent's goal and the informativity of their future actions. It then weighs the utility of social learning against the utility of self-exploration (non-social learning). Using a multi-player treasure hunt game, we show that our model can quantitatively capture human trade-offs between social and non-social learning. Furthermore, our results indicate that these two components allow agents to flexibly apply social learning to achieve their goals more efficiently.

\end{abstract}

\section{Introduction}

Social learning is a powerful mechanism through which humans and non-human animals acquire information about the world. It often requires observing other agents' actions and extracting information from these observations. However, the payoffs of social learning are not always immediately clear; non-social learning (e.g., through trial-and-error) can sometimes be just as (or even more) powerful and efficient. \blfootnote{Published as a conference paper at CogSci 2025}

Imagine you are deciding a restaurant to go for dinner. There are many restaurant options and you don't know which ones are good. You can go through their menus (nonsocial learning) and decide or observe which restaurants others tend to go to (social learning). However, because people have different food preferences, social learning may not lead to the most optimal outcome as many may be heading towards a famous Italian restaurant but you are craving for Asian food.




Much of the research on social learning has focused on various heuristics, such as copying the most successful agent or the majority \citep{heyes2012s, rendell2011cognitive, muthukrishna2016and, laland2004social}. While these heuristics can be powerful, and seem to be attested in behavioral data, they are probably not the whole story. Humans are able to use more sophisticated learning strategies based on reasoning about the beliefs and desires of other agents \citep{gershman2017learning,gweon2021inferential,hawkins2023flexible}. We refer to such strategies as ``theory-of-mind'' (ToM) social learning. These strategies enable learning from sparse, indirect data. For example, you might infer that your friend knows a good restaurant but isn't hungry, or is hungry but doesn't know a good restaurant; only when you can infer that both conditions are true should you follow your friend. You could also observe your knowledgeable friend when she's hungry so that you can use this information later when you're hungry. Or you could wait until your ignorant friend acquires knowledge before following them.


Most existing studies on social learning have mainly focused on tasks where all agents are pursuing the same goal. For example, in the collective sensing task used by \citet{hawkins2023flexible}, once the observer can infer who has knowledge of the goal location, social learning (observing knowledgeable agents) tends to be less costly than non-social learning (direct exploration in the environment). In more realistic settings, agents may have different goals, and direct exploration may sometimes be more effective. While some recent work has started to explore human social learning in settings with diverse agent goals \cite{witt2024humans}, they do not explicitly propose how humans estimate the utilities of social vs non-social learning strategies in complex tasks. 

Estimating the utility of social learning becomes challenging in a complex multi-agent domain where agents have diverging goals and knowledge. The benefit of observing others, despite their expertise, may not always outweigh the cost, because even if the agents are knowledgeable, they may not produce behaviors that are more useful to an observer than direct experience. Even in cases where we can infer useful information from these behaviors, social learning can quickly become expensive and time consuming if it requires observing long trajectories. Our aim is to understand how humans balance these trade-offs as they decide when to engage in social vs. non-social learning.

\begin{figure}
    \centering
    \includegraphics[width=0.95\linewidth]{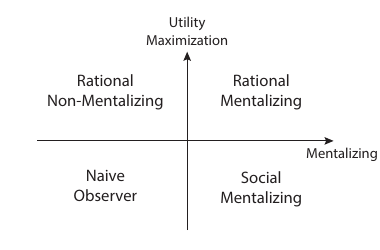}
    \caption{Two major components of the full rational mentalizing model. Mentalizing about other agents' goals and future actions allows the observer to estimate the utility of social learning. Utility maximization enables the observer to weigh the utility of social learning against non-social learning. Three alternative models occupy different parts of the model space (see text for details).}
    \label{fig:baselines}
\end{figure}

We propose a ``Rational Mentalizing'' model of social learning grounded in prior work on Bayesian ToM \citep{baker2017rational,jara2019theory}. The model estimates the utility of social learning by mentalizing about other agents' goals and plans, which it weighs against the utility of non-social learning to decide if and when to observe the other agent. We test the model in a complex multi-agent treasure hunt game where each player needs to find one hidden object in order to pass a barrier and reach their goal. On each step, the main agent, controlled by the participants, can either observe another agent for one step with a low cost, or perform an action with a higher cost. We created different trade-offs by varying the layout of the environment, allowing us to disentangle the predictions of the rational mentalizing model from several alternatives (Figure \ref{fig:baselines}).


\section{Related Work}

\subsection{Social learning and non-social learning}

Social learning and non-social learning are two dominate ways humans and non-human animals apply to acquire new information and skills. The former enables agents to observe others to gather social information while non-social learning primarily relies on reward learning by trial-and-error. Studies in evolutionary biology and animal cognition have long explored the "when" and "who" in social learning and suggested that social learning gives social learners evolutionary advantages as they can gather information more efficiently than non-social learners \citep{kendal2005trade, heyes2012s, kameda2003does, henrich2016secret, kendal2018social}. However, most existing studies treat social learning as copying or imitating other agent's actions or policies. Human social learning, on the other hand, is uniquely complex, flexible and efficient, as we don't simply copy successful agents, but use higher level cognitive skills to reason about the secrets behind their success \citep{horner2005causal, gweon2021inferential, bonawitz2016computational}.

\subsection{Theory of Mind}
Theory of Mind refers to human's ability to reason about other agents' mental states such as goals and beliefs. It's a critical cognitive skill for human social interaction \citep{wellman2004theory, wellman2002understanding}. Recent work in cognitive science has increasingly suggested that Theory of Mind plays a crucial role in human teaching and learning \citep{gweon2021inferential, shafto2012learning, hawkins2023flexible, dutemple2023know}. 

Our work builds on the Bayesian Theory of Mind framework by \citet{baker2017rational}, which models other agents as rational planners and infers their goals and beliefs by inverting a generative model of agent's goal-directed actions. The BToM model has been used for modeling a variety of human social behaviors that involve mentalizing \citep{zhixuan2024pragmatic, ying2024grounding, stacy2021modeling}. 

\subsection{Rational Utility Maximization}
Our work is also related to research on human rational decision making, which models humans as reward maximizers \citep{gershman2015computational, simon1955behavioral}. Past research in social learning has shown evidence of utility computation in humans' adaptive social and non-social learning strategies. For example, \citet{kendal2005trade} shows that people are more likely to engage in social learning when the cost of non-social learning increases. However, as far as we know, there has not been any studies on estimating the utility of social learning.

\section{Computational Model}

In this section, we describe our rational mentalizing model of social learning. We start by formulating the problem in a partially-observable multiagent setting. We then uses Bayesian Theory-of-Mind (BToM) as a computational framework for inferring agents' goals and beliefs, which we then use for utility estimation by simulating the other agents' future actions and evaluating their informativity for the observer's own planning. 

\subsection{Problem Formulation}

We formulate the setting as a two-player decentralized partially observable markov decision process (Dec-POMDP), described by the tuple $(\mathsf{S}, \mathsf{A}, \mathsf{C}, \mathsf{G}, \mathsf{\Omega}, \mathsf{O}, P_s, P_i)$, where $\mathsf{S}$ is the set of environment states, $\mathsf{A} = \mathsf{A}^m \times \mathsf{A}^o$ the set of player actions, $\mathsf{A}^{m} = \{a^m\}$ is the set of main agent actions, $\mathsf{A}^{o} = \{a^{o}\}$ is the set of other agent actions, $C: \mathsf{S} \times \mathsf{A} \to \mathbb{R}$ is a cost function that maps state-action transitions to real numbers, $\mathsf{G}$ is a set of possible goals $g^m$ for the main agent and $g^o$ for the other agent, each $g^m, g^o \subseteq S$ is a set of terminal states, $\mathsf{\Omega}$ is the set of observations the agents can make, $\mathsf{O}$ is the agents’ observation function $P(o|s)$, $P_s(s'|s, a^{m},a^{o})$ is the environment transition distribution, and $P_i(s_i, g)$ is a distribution over initial states of the game.

If an agent does not have full observability, their belief at timestep t about the state is $b_t$, which gets updated at each timestep.

\subsection{Modeling agent's planning and social inference}
Following prior work on rational planning and inverse planning \citep{zhixuan2024pragmatic, ying2023inferring}, we first define a forward model for planning by agent $i$:
\begin{alignat}{2}
\textit{Goal Prior:}& \quad P(g^i) \label{eq:goal-prior} \\
\textit{State Prior:}& \quad P(s_0) \label{eq:state-prior} \\
\textit{Belief Update:}& \quad b^i_t \sim P(b^i_t | o_t) \label{eq:belief-update} \\
\textit{Action Selection:}& \quad a^i_t \sim P(a^i_t | b^i_{t-1}, g^i) \label{eq:action-selection} \\
\textit{State Transition:} & \quad s_t \sim P(s_t | s_{t-1}, a^i_t) \label{eq:state-transition}
\end{alignat}
where
\begin{equation}
    P(a^i_t | b^i_{t-1}, g) = \frac{\exp\left(-\beta \hat Q_g(b^i_{t-1}, a^i)\right)}{\sum_{a'} \exp\left(-\beta \hat Q_g(b^i_{t-1}, a')\right)}
\end{equation}
is the agent's policy with temperature $\beta > 0$, which indicates action optimality. Here $\hat Q_g(b_{t-1}, a^i)$ represents the estimated cost of the optimal plan to achieve goal $g^i$ after taking action $a^i$ starting at the believed state $b^i_{t-1}$. 

Then given observations of the agent's actions for $T$ timesteps, the agent's goal and beliefs can be inferred by inverting the conditional probability:
\begin{multline}
    P(g^i, b^i_{0:T} | a^i_{1:T}) \propto \textstyle\sum_{s_{0:T}} P(g^i) P(s_0) P(b^i_0 | s_0) \\ \textstyle\prod_{t=1}^T P(a^i_t | b^i_{t-1}, g) P(s_t | s_{t-1}, a^i_t) P(b^i_t | s_t). \label{eq:inference}
\end{multline}

\subsection{Estimating the utility of social vs non-social learning}

We estimate the utility of the main player $m$ observing ($a^m = \text{Obs}$) the other player $o$ at timestep $t$ by simulating possible goals and plans the other agent may pursue, then estimating how the other agent's future goal-directed actions may be useful for the observer's own planning, weighted by the likelihood of the goal inferred from past action observations in Eq. \ref{eq:inference}:
\begin{align}
    U(\text{Obs},b^m_t, g^m) = \sum_{g^o \in G}P(g^o)U(\text{Obs}, b^m_t, g^m | g^o),
\end{align}
Each conditional utility function can then be calculated as the total cost of the agent's plan under social learning which includes observing the other agent for T steps then moving to retrieve the goal. We assume the agent's observing is optimal where they stop observing when the benefit of additional observations doesn't offset the observation costs.
\begin{align}
    &U(\text{Obs},b^m_t, g^m| g^o) = \sum_{b_t^o}P(b_t^o) \max_{T} {U(\text{Obs},b^m_t, g^m, T | g^o, b_t^o)} \nonumber\\
    &= \sum_{b_t^o}P(b_t^o)\max_{T} \left[ - T \times C(Obs) - \sum_{a^m \in \pi(g^m, b^{m\prime}_t)}C(a^m) \right]
\end{align}
where $T$ is the number of observe actions performed by the observer, $C$ is the cost function, and $\pi$ is the planner that returns a set of actions given the agent's goal and beliefs, and  $b^{m\prime}_t$ is the updated main agent's belief after observing the other agent's action for T steps $a^o_{1:t+T}$, which can be simulated by running forward planning with other agent's goal $g^o$ and possible belief states $b^o_t$.

On the other hand, the utility for acting rather than observing ($a^m = \text{Act}$) is:
\begin{align}
    U(\text{Act},b^m_t, g^m) = -\sum_{{a^m \in \pi(g^m, b^m_t)}} C(a^m),
\end{align}
which is the total plan cost under self-exploration. Then at each timestep t, the agent chooses to observe only if the utility of social outweighs the utility of non-social learning.
\begin{align}
    U(\text{Obs},b^m_t, g^m) > U(\text{Act},b^m_t, g^m).
\end{align}

In our implementation, we use an A star search planner \citep{astar} for simulating the agent's future actions.

\section{Experiment}

To evaluate our model on people's social learning behavior, we designed a multiagent game where the player can choose to either observe the other player or explore the maze by themselves. This game design enable us to investigate how humans flexibly use social learning and nonsocial learning to most efficiently achieve their goal and what factors influence people's social learning and nonsocial learning behavior. We collected data on human participants playing the game and analyzed their observing behaviors under carefully designed game levels.

\subsection{Domain and scenarios}

Our domain is a mutli-agent treasure hunt game, inspired by the single-agent door-keys-gems domain used in prior work \citep{ying2024grounding, zhi2020online} for studying agent's rational planning and theory of mind.

\begin{figure}
    \centering
    \includegraphics[width=\linewidth]{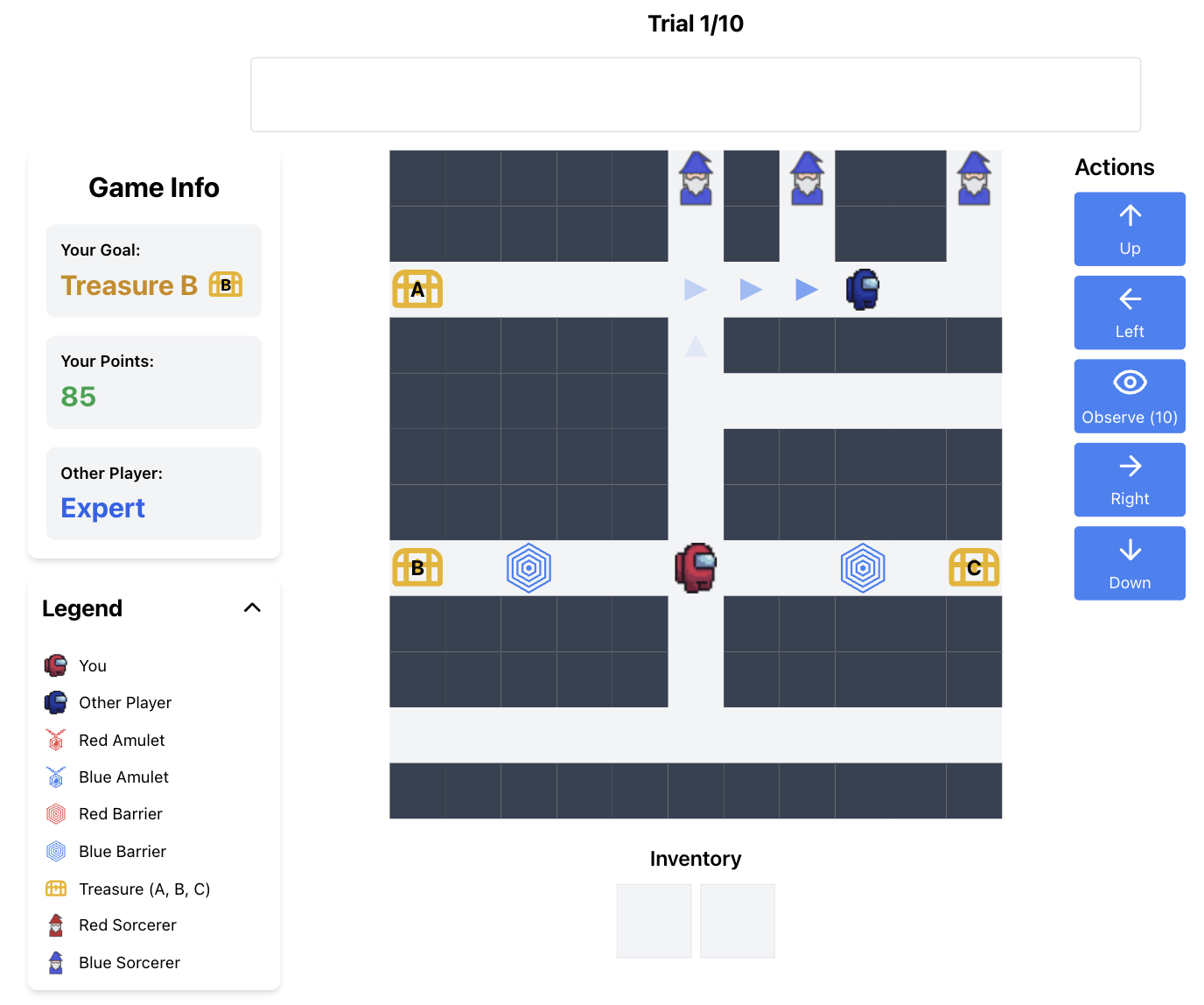}
    \caption{Experiment interface for the multiagent treasure hunt game. At each timestep, the main player in red can choose to either observe the other agent in blue or explore the maze themselves.}
    \label{fig:interface}
\end{figure}

The game interface is shown in Figure \ref{fig:interface}. In this game, there are 3 treasure chests labeled with letters A, B, C. These treasure chests are either unobstructed or blocked by a red or blue barrier. In order to pass through a barrier, the player needs to obtain an amulet of the same color from the wizards that reside in the map. The number of blue and red wizards varies across maps. Furthermore, among all blue (red) wizards, only one has a blue (red) amulet. Multiple agents can play the game concurrently. They can see each other's location, yet each agent's action does not affect any other agent's game state. In other words, any agent who gets an amulet from a wizard, passes through a barrier and retrieve a treasure box doesn't affect the other agent doing the same.

In our experimental setup, there are two agents playing the game. The red agent is the observer agent controlled by the human whereas the blue agent is an NPC being observed. The two agents each has a goal of retrieving one of the three gold chests, but they do not know the goal of the other agent. The observer agent's goal is given to the human player at the beginning of the trial. 

The red agent does not know which wizard has the amulet yet the red agent knows that the blue agent has full knowledge about the location of the amulet (Other agent's level = Expert) and therefore the red agent can observe the blue agent and see if they walk towards any blue wizards. At each timestep, the red agent can choose to either observe, in which case the other agent moves one step, or move in one of the four directions (up, down, left, right), in which case the other agent stays still. 

We constructed 27 game maps, each with two variants with a total of 54 stimuli. Each variant of the map corresponds to different goal and path pair of the other agent, where in one stimulus the other agent has the same goal as the observer agent and in the other stimulus their goals differ. Across the stimuli we vary goals and locations of the main agent and the other agent in order to test participants' social learning behavior across a variety of scenarios. 



\begin{figure*}[ht!]
    \centering
    \includegraphics[width=1\linewidth]{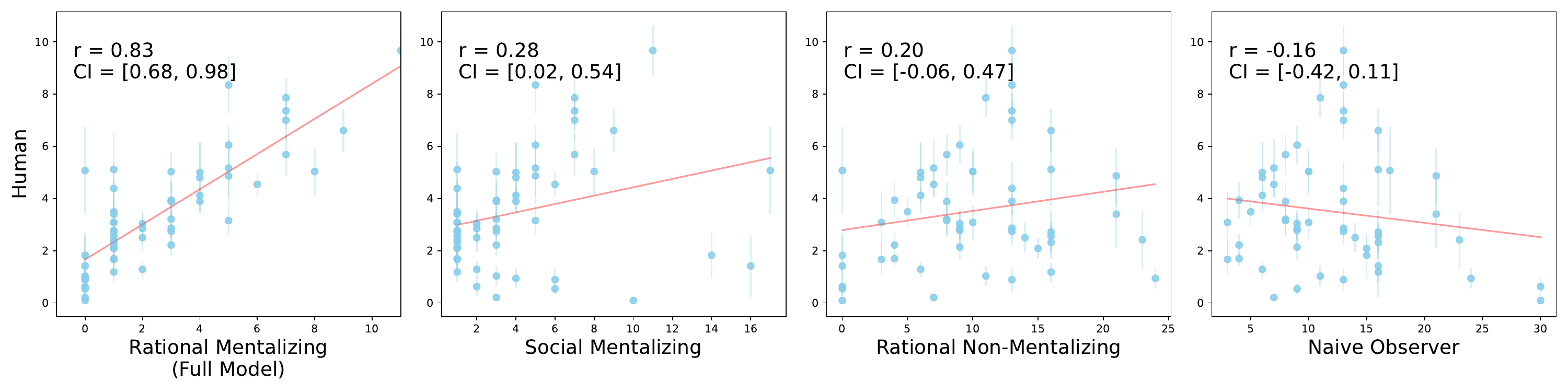}
    \caption{Correlation plot comparing model predictions and human behavior. Each point represents average model or human social learning behavior (the number of observe steps) on a given game level. Error bars indicate standard error. CI indicates 95\% confidence interval.}
    \label{fig:scatter}
\end{figure*}

\subsection{Experimental Procedure}

Our study was conducted through an online interface. The participants were first presented with instructions and then an interactive tutorial. Each participant then completed 10 different trials, each containing a different map. 

In each trial, the participants were given a fixed amount of points at start, which were spent as they perform different actions. The participants received the leftover points after they completed the assigned goal in each trial. At the end of the experiment, the participants received a bonus payment for the points they accumulated.

\subsection{Human Participants}
We recruited 222 US participants through Prolific (mean age = 37.32, 119 female, 99 male, 2 non-binary, 2 others). The participants were paid 15 dollars per hour. We excluded 16 participants who had a final score below 0.

\begin{figure*}[ht!]
    \centering
    \includegraphics[width=0.92\linewidth]{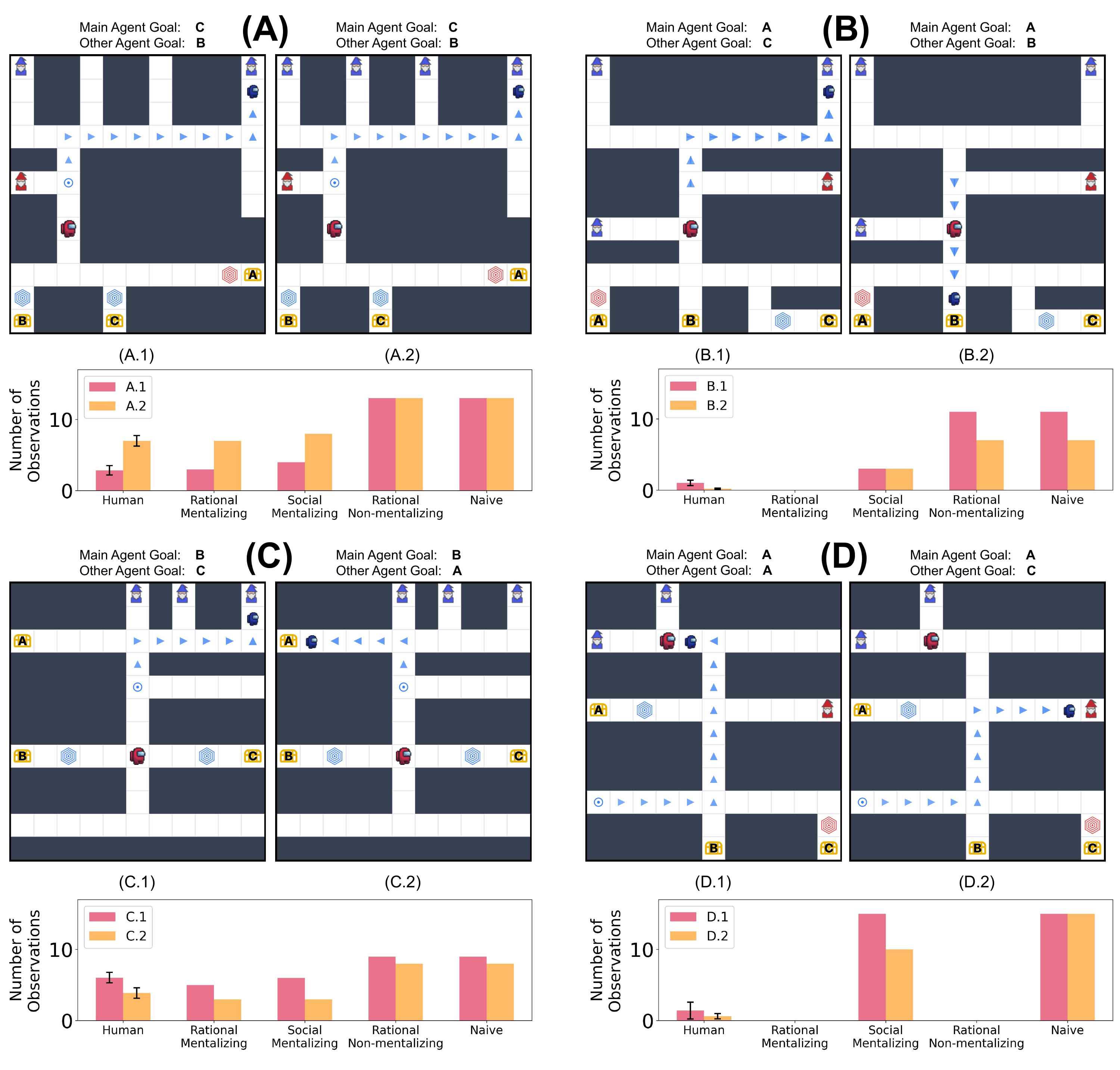}
    \caption{Four qualitative examples comparing human and model observing behaviors. In examples A and B, we show that the number of observations taken by the Rational Mentalizing model and human participants is sensitive to the quantity and placement of the blue wizards. In example C we show mentalizing about the other agent's goal is essential in determining whether the agent should observe the other agent. In example D we show that humans and the Rational Mentalizing model can rationally decide not to observe the other agent when non-social learning is cheaper than social learning. Error bars indicate the standard error.}
    \label{fig:qualitative}
\end{figure*}

\subsection{Baselines}

Our model has two key components: mentalizing about other agents' mental states from observations and rational decision making by comparing expected utilities for social learning and non-social learning. To evaluate the criticality of the two components, we include three ablation models as baselines by removing one or both of the components. The number of observations is capped at 15 steps for each trial.
\begin{itemize}
    \item Naive Observer: The agent always observes until the other agent stops moving (reaches their goal) or interacts with a wizard. Once the other agent interacts with a wizard, the player knows that wizard possess an amulet.
    \item Rational Non-mentalizing Observer: The agent compares the utility of self-exploration against social learning to decide whether to observe the other agent. However, the observer does not infer the other agent's mental states. The observer always assumes the other agent has the same goal and observes until the other agent stops moving or interacts with a wizard. 
    \item Social Mentalizing Observer: The agent does not evaluate the utility of nonsocial learning. The agent decides to observe when they expect the other agent's future actions to be informative about the environment state:
    \begin{align}
        \max_{T} \,\textrm{KL} [P(b^m_t) ||P(b^m_t|a^o_{t:t+T})] > 0
    \end{align}

\end{itemize}

\section{Results}

\subsection{Quantitative analysis}

As shown in Figure \ref{fig:scatter}, the Rational Mentalizing model correlated strongly with human behavior, with $r = 0.83$, where the split-half correlation among human participants is 0.8. In contrast, the baseline models fitted poorly. We also compared the average number of steps and planning cost to complete each trial for humans and models in Table \ref{tab:summary}. The Rational Mentalizing model closely matched the human statistics, although humans on average tend to observe slightly more than the Rational Mentalizing model (still much less than the non-mentalizing models). This might be because human subjects assume that the other agent's last action could be due to a mistake, and therefore they often observe 1 or 2 steps more to accumulate additional evidence that the other agent is indeed walking towards a certain wizard or chest. The Social Mentalizing model also matched the results fairly well, though not quite as closely. These results support the claim that mentalizing about other agents and rational utility maximization are both indispensable for explaining flexible human social learning.

\begin{table}[ht!]
    \centering
    \begin{tabular}{lcc}
        \toprule
        & \textbf{Average Steps} & \textbf{Average Cost} \\
        \midrule
        Naive Observer        &  46.43 (2.03)&86.02 (4.27)  \\
        Rational Non-mentalizing &  44.20 (2.41)& 83.68 (4.68)  \\
        Social Mentalizing &  39.35 (1.75) & 78.93 (4.03)\\
        Rational Mentalizing  &  \textbf{37.24 (2.06)} & \textbf{76.72 (4.37)}  \\
        Humans                 &  37.89 (2.00) & 77.43 (4.06)  \\
        \bottomrule
    \end{tabular}
    \caption{Average model and human performances on the treasure hunt game. The rational mentalizing model took a similar number of steps as did humans and is the most efficient among all models. Standard errors are shown in parentheses.}
    \label{tab:summary}
\end{table}

\subsection{Qualitative analysis}

To lend further insights into how mentalizing and utility maximization work together to enable flexible social learning, we highlight some diagnostic qualitative examples in Fig. \ref{fig:qualitative}.

In example A, we show how small modification of the map can affect both human and model observing behaviors, where A.1 and A.2 have the same map but different numbers of blue wizards. In this example, the main agent's goal is to get Treasure Chest B. By mentalizing over the other agent's goal, both mentalizing models reason that the other agent's goal is either B or C, both of which require getting a blue amulet from one of the wizards. 



In A.1, human participants observed 2.86 steps on average, similar to the mentalizing models (around 3 steps). This is because at timestep 3, when the other agent turned right, the mentalizing observer can infer that the blue amulet is with the blue wizard on the top right. In A.2, when two more wizards were added, the number of observation steps increased to 7.0 for both human participants and the mentalizing models, where the other agent walked past the third blue wizard from left. These examples show that mentalizing models can flexibly adjust its social learning behavior to adapt to different different contexts through using theory of mind.

In example B, the main agent's goal is to retrieve Treasure Chest A, which requires a red key from the red wizard. Most human participants, as well as the rational mentalizing model, chose not to observe the other agent. The social mentalizing model, on the other hand, observed the other agent until it can infer which blue wizard has the amulet (B.1) or the other agent is heading towards Treasure Chest B. This example shows that the utility of social learning hinges on the goal of the learner; humans actively seek \textbf{goal-relevant} information through social learning.

In example C, the other agent's goals and plans necessitate different instrumental actions. In C.1, where the other agent's path is informative, humans and the mentalizing models observed 6.05 and 5 steps, respectively, whereas the non-mentalizing models observed 9 steps. On the other hand, in C.2, both humans and the mentalizing models stopped observing at timestep 3.88 and 3, respectively, when they realized the other agent is heading towards treasure A and the future actions would not be informative for the observer's own goal and plan. A non-mentalizing observer would need to observe until the other agent interacts with the blue wizard or stops moving after retrieving the goal chest. These results show that mentalizing enables agents to infer other agents' latent knowledge of the world without observing the full trajectory.

In example D, we show how rational utility maximization enables the observer to decide whether to engage in social learning. The other agent is situated far away from the two blue wizards, while the main agent is close to them. In this case, non-social learning is cheap while social learning can be costly, as the observer is not only uncertain about the other agent's goal, but also needs to observe for many steps in order to recover any social information. As a result, most human participants and the rational mentalizing model chose not to observe the other agent. The social mentalizing model, on the other hand, chose to observe until it can infer which blue agent has the amulet (D.1) or realizes that the other agent is heading towards a red wizard (D.2).

\section{Discussion and future directions}

In this paper, we proposed a rational mentalizing model of human social learning, identifying two key secrets to the success of flexible and efficient human social learning: (1) the ability to reason about other agents' goals, beliefs and plans; and (2) the ability to flexibly switch between social and non-social learning by estimating their relative utility. Using a multiagent treasure hunt game, where the main player can gather information by either observing the other agent or exploring by themselves, we found that the rational mentalizing model matches human observing behaviors with quantitative accuracy, whereas alternative models fitted poorly and were less efficient in recovering the relevant information.

Our work has several limitations. First, our game setup is quite simple. In the real world, the decision to engage in social vs non-social learning may be influenced by other contextual and idiosyncratic factors, such as an individual's general propensity to observe (sociality vs non-sociality). Future work can enrich the rational mentalizing model to more fully capture the range of human social learning behaviors.

Second, in our setup the observed agent has full knowledge of the maze and object placements. This is often not true in human social learning tasks. For our next step, we aim to model if and how humans learn from others who may not have the perfect information.

Third, humans often learn in a complex social environment with the presence of multiple agents or groups. We only included one expert agent in our experimental setup. Future work can scale the rational mentalizing model to model social learning behavior in crowds of players with diverging interests, expertise, and goals.


\printbibliography

\end{document}